\documentclass[twocolumn,showpacs,preprintnumbers,amssymb]{revtex4}
\usepackage{pstricks}%
\usepackage{graphicx}
\usepackage{dcolumn}
\usepackage{bm}
\usepackage{dsfont}
\usepackage{amsmath}
\usepackage{pifont}
\usepackage{psfrag}
\usepackage[latin2]{inputenc}

\newcommand\bra[1]{\langle{#1}|}
\newcommand\ket[1]{|{#1}\rangle}

\newcommand{\BE}{\begin{equation}}
\newcommand{\EE}{\end{equation}}

\newcommand{\skipc}[2]{}

\newcommand{\fig}[1]{Fig.~\ref{#1}}
\newcommand{\eq}[1]{Eq.~(\ref{#1})}
\newcommand{\Sec}[1]{Sec.~\ref{#1}}

\newcommand{\I}{\ensuremath{{\mkern1mu\mathrm{i}\mkern1mu}}}
\newcommand{\E}{\ensuremath{{\mkern1mu\mathrm{e}\mkern1mu}}}

\begin{document}

\preprint{APS/123-QED}

\title{Accurate phase measurement with classical light}

\author{Sabine W\"olk}
\author{Wenchao Ge}
\author{M. Suhail Zubairy} %

\affiliation{Institute of Quantum Science and Engineering (IQSE) and Department of Physics and Astronomy\\
Texas A$\&$M University, College Station, TX 77843-4242}%

\date{\today}

\begin{abstract}
In this paper we  investigate whether it is  in general possible
to substitute maximally path-entangled states, namely NOON-states
by classical light in a Doppleron-type resonant multiphoton
detection processes by studying  adaptive phase measurement with
classical light. We show that multiphoton detection probability
using classical light coincides with that of NOON-states and the
multiphoton absorbtion rate is not hindered by the spatially
unconstrained photons of the classical light in our scheme. We
prove that the optimal phase variance with classical light can be
achieved and scales the same as that using NOON-states.
\end{abstract}


\maketitle

\section{Introduction\label{intro}}
Optical phase measurement is the basis of many scientific research
areas, such as quantum metrology and quantum computing. The
precision of optical phase measurement is bounded by the standard
quantum limit (SQL) or shot noise limit (SNL) which scale as
$1/\sqrt{N}$ in the number of $N$ independent resources. However,
many authors \cite{Caves1981, Yurke1986, Sanders1995} proposed the
possibility to beat the SQL and reach the Heisenberg limit which
scales as $1/N$ by using nonclassical states. One possibility to
reach the Heisenberg limit is to use NOON states and combine them
with an adaptive measurement scheme \cite{Berry2009}

NOON states are among the most
highly-entangled states and they have the potential to enhance
measurement precision \cite{Lee2002}  not only in phase measurement \cite{Berry2009} but also in subwavelength lithography
\cite{Boto2000} and atomic interferometry \cite{Wei2011}.

 However, due to the
difficulty in generating NOON states with large photon number
($N>2$), alternative efforts have been made such as by the use of
multiple passes of single photon \cite{Higgins2007} and dual Fock
state \cite{Xiang2010}. In the single photon scheme, the
measurement time scales with $N$ which poses a problem for very
fast measurement. In the dual Fock state scheme, a sub-SNL rather
than Heisenberg limit is obtained. Moreover, number entangled
states \cite{Boto2000} are criticized for being highly spatially
unconstrained to be absorbed at a tiny spot
\cite{Steuernagel2004}.

Recently, Hemmer \emph{et al.} \cite{Hemmer2006} pointed out the
quantum feature of path-number entanglement with NOON states can
be realized with classical light. Their work shows the possibility
of highly frequency selective Doppleron-type multiphoton
absorption process \cite{Haroche1972,Kyrola1977,Berman1977} with
classical light used to achieve subwavelength diffraction and
imaging. The multiphoton absorption process generates and detects
a NOON state simultaneously. We apply this semiclassical
frequency-selective measurement with classical light in the
subwavelength lithography to optical phase measurement. Our
proposal provides an effective alternative adaptive phase
measurement method in conventional two-path interferometry, which
previously required nonclassical states. Although a point detector
is also required in our scheme, enough number of photons from the
classical light will arrive at the point of detector to stimulate
frequency-selective measurement. Photons arrive at other positions
are of no interest to us. Therefore, the excitation rate in our
scheme is not hindered by spatially unconstrained photons of the
classical light.

In this paper, we first summarize in \Sec{sec:berry} the main
ingredients of  phase measurement with NOON
states~\cite{Berry2009}. Then we apply in \Sec{sec:substitution}
the idea of replacing NOON-states by classical light to this phase
measurement algorithm  before we discuss the detection rate
scaling and possible errors of our algorithm in \Sec{sec:scaling}

\section{Phase measurement scheme using NOON states \label{sec:berry}}

\begin{figure}
 \includegraphics[width=0.45\textwidth]{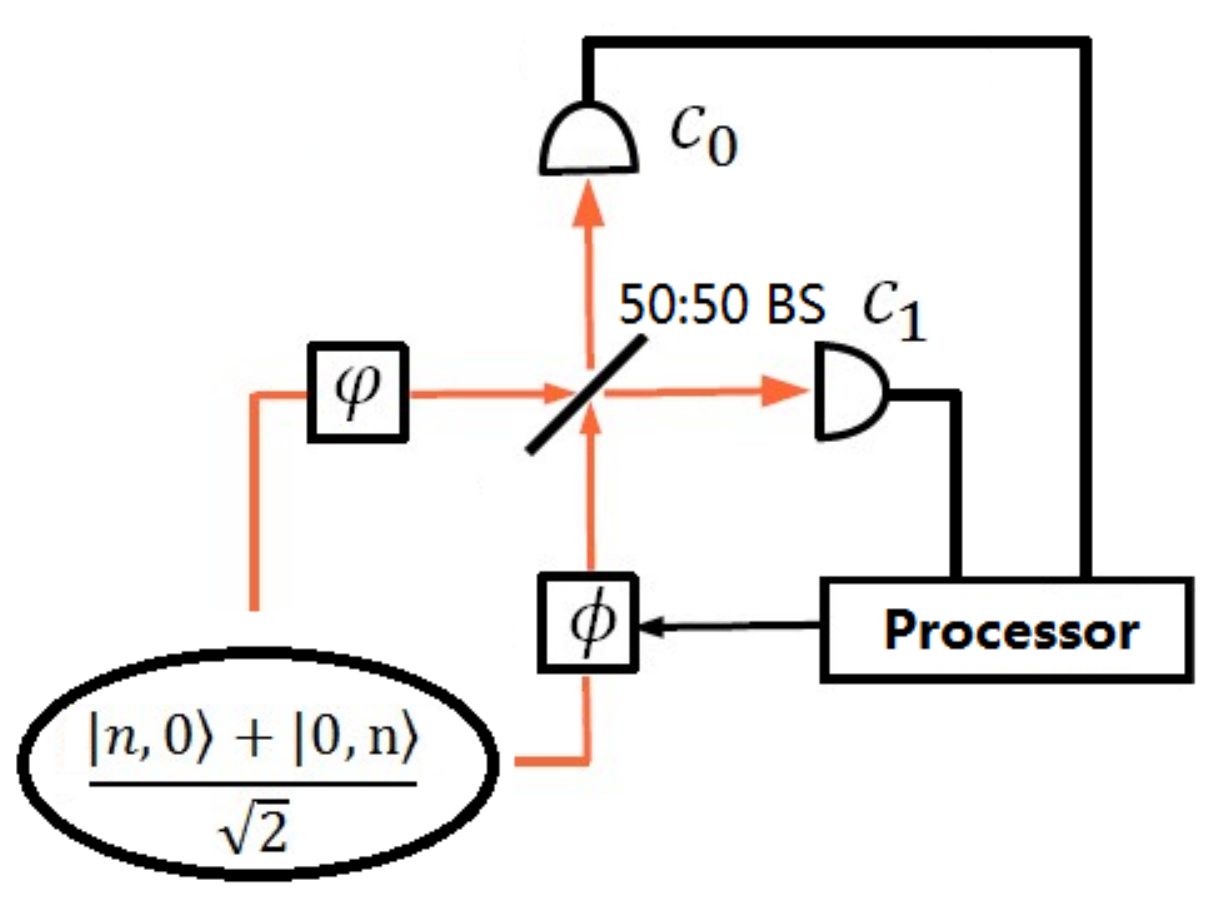}
\caption{Measurement scheme for the unknown phase $\varphi$ with the help of the NOON state $\left(\ket{n,0}+\ket{0,n}\right)/\sqrt{2}$\label{fig:phase_meas}}
\end{figure}

To perform a phase measurement Berry et.~al. \cite{Berry2009} use
a sequence of NOON states $(\ket{n,0}+\ket{0,n})/\sqrt{2}$ with
$n= 2^k$, where $k$ is varied from $K$ to $0$. These states are
sent through a Mach-Zehnder interferometer  as shown in
\fig{fig:phase_meas}, where the phase of the  light in one arm is
shifted by the unknown phase $\varphi$ and by a controllable phase
$\phi$ in the other arm. The photons are detected in the two
output modes $c_0$ and $c_1$ after the 50/50 beam-splitter. The
detection of all photons, with measurement results $\vec{u}_n=
\left\{u_1,u_2,\dots, u_n\right\}$ and $u_j \in \{0,1\}$ obeys the
probability distribution \BE P(\vec{u}_n|\varphi)\equiv
\frac{1}{2}\left\{1+(-1)^u \cos[n (\varphi-\phi)]\right\}
\label{eq:P(u,phi)} \EE where $u$ is given by the parity (even or
odd) of $u_1+u_2+ \dots + u_n$. In order to explain the phase
measurement scheme, we first assume that the unknown phase
$\varphi$ is of the form \BE \varphi\equiv\varphi_K \equiv
\pi\sum\limits_{k=0}^{K}\frac{a_k}{2^k} \EE with $a_k \in
\{0,1\}$. We start with $n=2^K$ and $\phi=0$, which leads to the
probability distribution \BE P(\vec{u}_n|\varphi_K)=
\frac{1}{2}\left\{1+(-1)^u \cos[(\pi a_K)]\right\}
\label{eq:P(u,phik)}. \EE which is equal to zero or unity
depending on $a_K$ and $u$ as specified below \BE
\begin{array}{c|c|c}
 &u=even&u=odd \\ \hline  a_K=0&1&0 \\a_K=1&0&1
\end{array}
\EE As a consequence, if we have measured $u$ equals an even
number, we know that $a_K$ must be equal to zero, because for
$a_K=1$, the probability to measure $u=even$ is zero. On the other
hand, if we have measured $u$ equals odd, then we know $a_K=1$.

For the next measurement, we choose $n=2^{K-1}$ and $\phi= \pi a_K/2^K$ and get $a_{K-1}$ similar to $a_K$, and so on until we have measured all coefficients $a_k$.

However, in general the unknown phase $\varphi$  consist  of an infinite number of coefficients and therefore, we can not measure them exactly. To determine the accuracy of this phase measurement algorithm, we discribe all measurements for different photon numbers $n$ by one POVM, given by
\BE F(\hat \varphi)\equiv
\ket{\varphi}\bra{\varphi}, \quad \ket{\varphi}\equiv
\frac{1}{N_K}\sum\limits_{j=0}^{N_K}\E^{\I j \hat \varphi}\ket{j}.
\EE
performed on the state
\BE \ket{\psi}\equiv
\frac{1}{(N_K+1)^{1/2}}\sum\limits_{j=0}^{N_K} \E^{\I j
\varphi}\ket{j} \EE
with $N_K=2^{K+1}-1$.

The scaling of the algorithm depends on the phase variance
$(\Delta \varphi)^2$ which is usually given by Holevo variance
$V_H\equiv\mu^{-2}-1$ with $\mu\equiv|\langle
e^{i\hat{\varphi}}\rangle|$. Thus the feedback phase $\phi$ should
maximize $\mu$ in the system phase ($\varphi$) probability
distribution. From Bayes' theorem, the probability distribution
for the system phase is then $P(\varphi|u)\propto
P(\vec{u}_n|\varphi)$ provided that $\varphi$ is an initially
completely unknown phase. Therefore, the sharpness of the phase
distribution in the semiclassical case is given by
\cite{Berry2009}
\begin{eqnarray}
\mu=\frac{1}{2\pi}\sum_{\vec{u}_n}|\int
e^{i\varphi}P(\vec{u}_n|\varphi)d\varphi|.
\end{eqnarray}
With maximized $\mu$ in the feedback process, the variance $V_H$
scales like the standard quantum limit for the measurement scheme
described above. However, by using $M$ copies of each NOON states
with $M\ge4$ and performing $M$ measurement for each NOON state,
Berry et.~al. show that this modified algorithm scales as the
Heisenberg limit \cite{Berry2009}. This algorithm was performed
experimentally \cite{Higgins2007} by using a single photon with
$n$ passes through the phase shift and for $M=6$ repeated
measurements for each $n$ given the system phase distribution Eq.
\eqref{eq:P(u,phi)}.

The main ingredients of this scheme are the enhancement of the
unknown phase $\varphi$ by the factor $n$ through NOON states and
the interference of the unknown phase $\varphi$ with the
controllable phase $\phi$. Furthermore, there are only two
possibilities for $u$: zero or one, and all probabilities
$P(\vec{u}_n|\varphi)$ add up to unity. Therefore, by knowing
$P(\vec{u}_n|\varphi)$ for one given $\vec{u}_n$, we can calculate
all other probabilities.

\section{Substitution of NOON states\label{sec:substitution}}

\begin{figure}
 \includegraphics[width=0.45\textwidth]{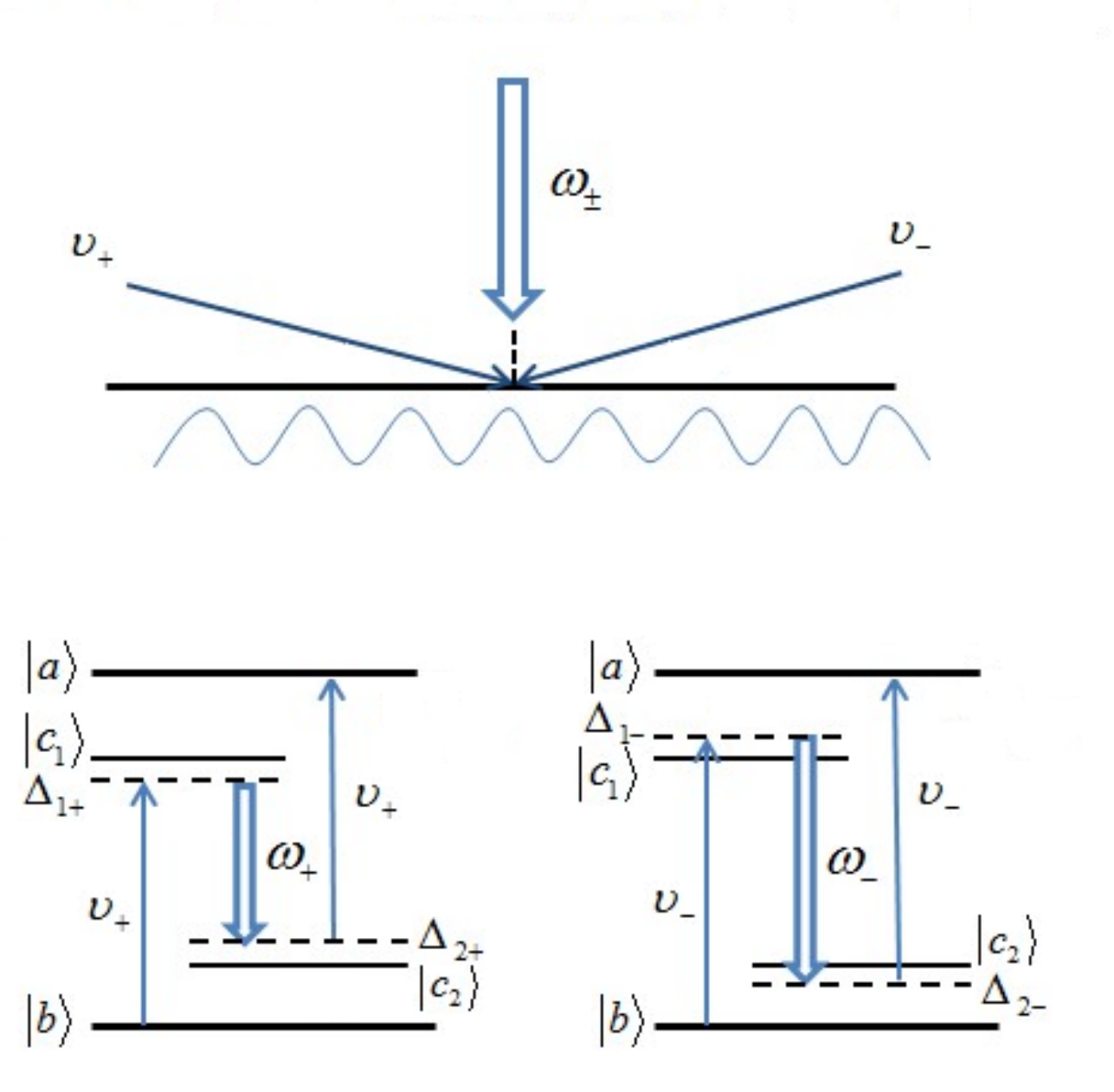}
\caption{Two photon interference with classical light and the
interaction of classical fields with the four-level atomic
structure. \label{fig:four-level scheme}}
\end{figure}

Let us now analyze the substitution of NOON states by classical
light as was done in \cite{Hemmer2006}. We first consider a
four-level atom as our point detector interacting with the
classical fields illustrated in Fig.~\ref{fig:four-level scheme}.
In this scheme, level $\ket{b}$ is assumed to be the ground level.
The intermediate levels $\ket{c_{1}}$ and $\ket{c_{2}}$ are highly
detuned from the driving fields and we do not include the
population decay rate from these levels. The population decay from
the upper level $\ket{a}$ is denoted by $\gamma$. We assume that
$\ket{b} \rightleftharpoons \ket{c_{1}}$,
$\ket{c_{1}}\rightleftharpoons \ket{c_{2}}$ and
$\ket{c_{2}}\rightleftharpoons \ket{a}$ are the only dipole
allowed transitions.

The basic idea behind this scheme is to send two signal beams of
slightly different frequencies $\nu_{\pm}$ from opposite
directions and two vertically incident driving beams of
frequencies $\omega_{\pm}$ as shown in Fig.~\ref{fig:four-level
scheme}. The excitation from level $\ket{a}$ to $\ket{b}$ can take
place by either absorbing two $\nu_{+}$ signal photons and
emitting one $\omega_{+}$ driving photon or absorbing two
$\nu_{-}$ signal photons and emitting one $\omega_{-}$ driving
photon. Any other process (such as absorbing one $\nu_{+}$ signal
photon and one $\nu_{-}$ signal photon and emitting one
$\omega_{+}$ driving photon or absorbing two $\nu_{-}$ signal
photons and emitting one $\omega_{+}$ driving photon) would be
forbidden by selection rules or be non-resonant and therefore
negligible for reasonable Rabi frequencies. These two excitation
branches then stimulate the $(\ket{2,0}+\ket{0,2})/\sqrt{2}$ NOON
state path correlations which is possible for super-sensitive
phase measurement when we shift the phases in each branch.

The two classical signal fields and one driven field interacting
with the four-level atomic system are written as \BE
E_S^+(x,t)\equiv\mathcal{E}_S \E^{\I (k_+x-\nu_+ t)}, \EE

\BE E_S^-(x,t)\equiv \mathcal{E}_S  \E^{\I (k_-x-\nu_- t)} \EE and
 \BE E_D\equiv \mathcal{E}_D
\left[\E^{-\I \omega_+ t}+\E^{-\I \omega_- t}\right]. \EE The
interaction Hamiltonian in the rotating wave approximation is
given by \BE
\begin{split}
H_I\equiv& \hbar \Omega_S
\left(\ket{c_1}\bra{b}\E^{\I\Delta_{1\pm}t+\I
k_{\pm}x}+\ket{a}\bra{c_2}\E^{\I\Delta_{2\pm}t+\I
k_{\pm}x}\right)\\&+\hbar
\Omega_D\ket{c_1}\bra{c_2}\E^{\I(\Delta_{1\pm}+\Delta_{2\pm})t}+\textrm{H.c.}.
\label{eq: HI}
\end{split}
\EE
The one-photon detunings given by $\Delta_{1\pm}\equiv
\omega_{c_1 b}-\nu_\pm$ and $\Delta_{2\pm}\equiv \omega_{a
c_2}-\nu_\pm$ are much larger compared with $1/t$ so that no atom
will be excited to the intermediate levels. The three-photon
resonance condition, $\omega_{ab}+\omega_{\pm}=2\nu_{\pm}$, is
considered when deriving the interaction Hamiltonian.

We consider the atomic system as a narrow bandwidth detector
\cite{Hemmer2006} for which the lifetime $1/\gamma$ in the excited
level $\ket{a}$ is longer than the detecting time $t$. In the
perturbative regime, $|\Delta_{j\pm}t|\gg1$, the amplitude of
excitation from $\ket{b}$ to $\ket{a}$ is given to the lowest
order by the three-photon process \cite{Scully2007}:
\begin{widetext}
\begin{eqnarray}
a^{(1)}(x,t) &=& (-\frac{\I}{\hbar})^3 \int_0^t dt_{1}
\int_{0}^{t_{1}} dt_{2}\int_{0}^{t_{2}}dt_{3} \bra{a}
\mathnormal{H_{I}(x,t_{1})}
\mathnormal{H_{I}(x,t_{2})}\mathnormal{H_{I}(x,t_{3})} \ket{b}
\nonumber \\
&=& \I\Omega_{S}^{2} \Omega_{D}t\left[ \left(\frac{e^{\I
2k_{+}x}}{\Delta_{1+}\Delta_{2+}}+\frac{e^{\I2
k_{-}x}}{\Delta_{1-}\Delta_{2-}}\right)+\left(\frac{(e^{-\I2\delta
t}-1)e^{\I 2k_{+}x}}{\Delta_{1+}(\Delta_{2+}+2\delta)(-\I2\delta
t)}+\frac{(e^{\I2\delta t}-1)e^{\I2
k_{-}x}}{\Delta_{1-}(\Delta_{2-}-2\delta)(\I2\delta t)}\right)\right.\nonumber \\
&+&\left. \left(\frac{e^{\I\delta
t}-1}{\Delta_{1-}(\Delta_{2-}-2\delta)(\I\delta t)} +
\frac{e^{\I\delta t}-1}{\Delta_{1+}\Delta_{2+}(\I\delta t)}
+\frac{e^{-\I\delta t}-1}{\Delta_{1-}\Delta_{2-}(-\I\delta t)} +
\frac{e^{-\I\delta
t}-1}{\Delta_{1+}(\Delta_{2+}+2\delta)(-\I\delta t)}\right)e^{\I
(k_{+}+k_{-})x}\right], \label{eq:first-order exp}
\end{eqnarray}
\end{widetext}
where $\delta\equiv\nu_{+}-\nu_{-}=\Delta_{j-}-\Delta_{j+}$
represents the frequency difference between the two signal beams.
The physical meaning of this amplitude of excitation is the
following: the first term in the square brackets represents
resonant three-photon process from each channel; the second and
the third terms are off resonant terms by interchanging one
driving photon $\omega_{\pm}$ and one signal photon $\nu_{\pm}$
respectively from the resonant three-photon process between the
two channels. Each of the non-resonant terms are multiplied by
either $\frac{e^{\pm\I \delta t}-1}{\pm\I \delta t}$ or
$\frac{e^{\pm\I 2\delta t}-1}{\pm\I 2\delta t}$. For $\delta
t\gg1$, 
\BE|\frac{e^{\pm\I j\delta t}-1}{\pm\I j\delta
t}|\leq \frac{2}{j\delta t}\ll1\quad (j=1,2),
\EE 
the contribution from the non-resonant terms is in general
proportional to $1/(\delta t/2)$.

 Under
the condition $\delta t\gg1$ and $\gamma t<1$, the only two
significant terms in the first order perturbation theory will be
the resonant term for which the same beam, $+$ or $-$, contributes
twice. Therefore, the amplitude of excitation $a^{(1)}(x,t)$ is
given by \BE a^{(1)}(x,t)\equiv \I \Omega_S^2\Omega_D t
\left(\frac{\E^{\I2k_+x}}{\Delta_{1+}\Delta_{2+}}+\frac{\E^{\I2k_-x}}{\Delta_{1-}\Delta_{2-}}\right).\label{eq:first-order}
\EE For $\Delta_{1+}\Delta_{2+} \approx
\Delta_{1-}\Delta_{2-}\equiv \Delta_{1}\Delta_{2}$ the probability
$P_a(x,t)$ to find the atom in the state $\ket{a}$ is given by \BE
P_a(x,t)= 2\left|\frac{\Omega_S^2 \Omega_D
t}{\Delta_1\Delta_2}\right|^2 \left[1+\cos(2k_+x-2k_-x)\right]. \EE
Note that this perturbation theory is only valid when the
effective Rabi frequency $\frac{\Omega_S^2 \Omega_D
}{\Delta_1\Delta_2}\ll1/t$ \cite{Hemmer2006}. This probability
shows an increased resolution by absorbing two photons each time
from each channel.

\begin{figure}
 \includegraphics[width=0.4\textwidth]{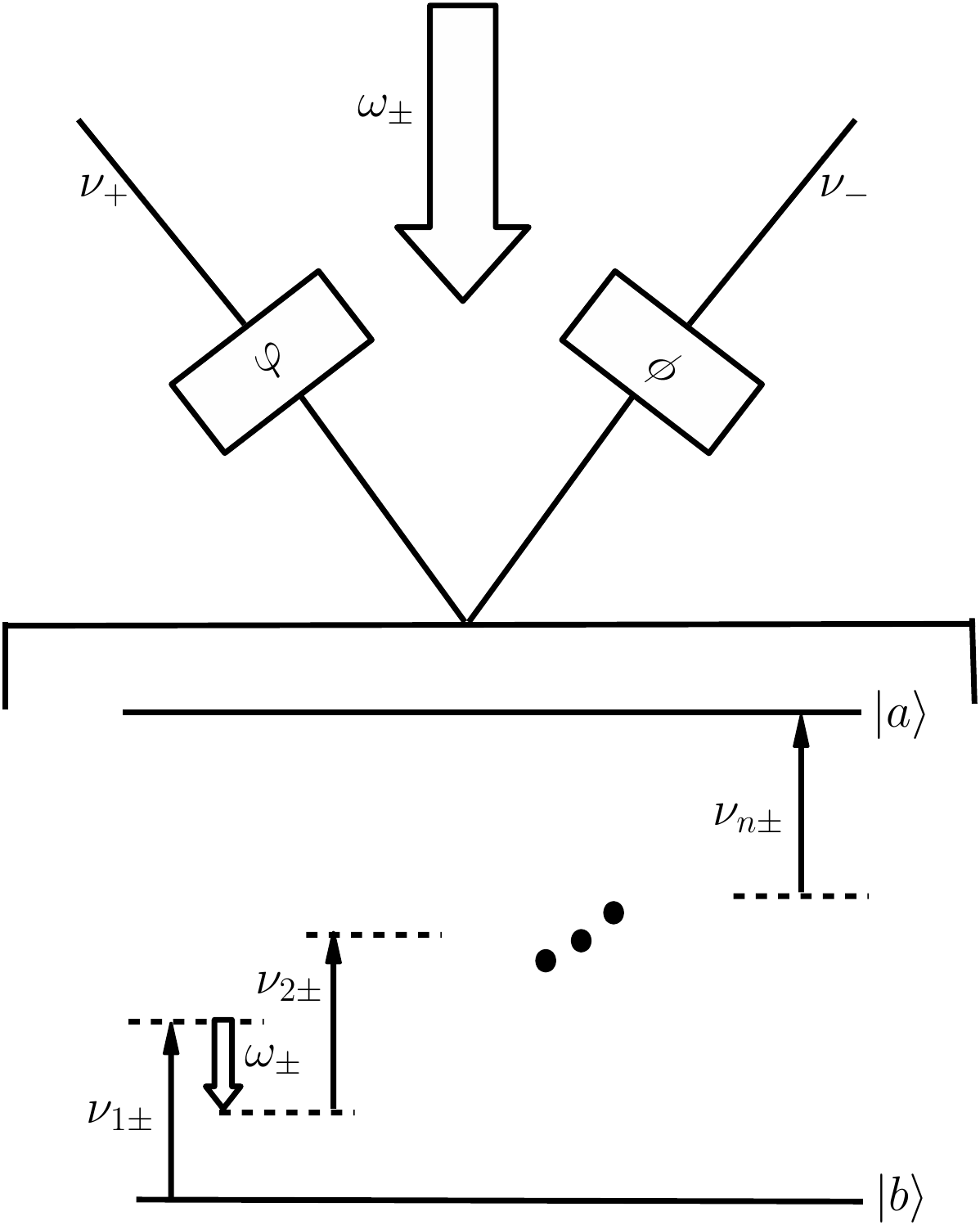}
\caption{The two signal fields with frequencies $\nu_+$ and
$\nu_-$ are shifted by the unknown phase $\varphi$ and the
controllable phase $\phi$, respectively. The drive fields
$\omega_\pm$ assist a directional resonance of $n$ photon
absorption.\label{fig:scheme}}
\end{figure}

This scheme can be generalized to an atom with $n$ intermediate
levels, where the two signal fields  are replaced by two bunches
of signal fields, which obey the following resonance condition \BE
\omega_{ab}=\sum\limits_{j=0}^{n}\nu_{j\pm}-(n-1)\omega_{\pm}. \EE

As explained above, under the condition $\delta t\gg1$ and $\gamma
t<1$, any interchange of photons between the two excitation
branches will result in a loss of resonance and any non-resonant
processes can be neglected. This relation ensures that the atom
absorbs $n$ photons from one branch of the signal beams ($\nu_{+}$
or $\nu_{-}$) and emits $(n-1)$ photons of the frequency
$\omega_+$ or $\omega_-$. By applying a phase shift $\varphi$ on
one of the signal fields and by $\phi$ on the other as illustrated
in \fig{fig:scheme} and assuming that $(k_+-k_-)x = 0$,  we obtain
\BE P_a(t|\varphi,\phi) = 2 \Omega_{eff}^2
t^2\left\{1+\cos\left[n(\varphi-\phi)\right]\right\}\label{eq:excitaiton_rate}
\EE with \BE \Omega_{eff}\equiv
\frac{\Omega_S^n\Omega_D^{n-1}}{\prod\limits_{j=1}^{n-1}\Delta_{2j-1}\Delta_{2j}},
\EE and the generalized perturbative regime is
$\Omega_{eff}t\ll1$.
 Therefore, the probability distribution $P_a$ behaves the same
way as the probability distribution \eq{eq:P(u,phi)} of a NOON
state used for measuring the unknown phase $\varphi$, but only
with the factor $u$ being fixed to an even number. However, if we
know the result of the function \BE 1+(-1)^u \cos[n
(\varphi-\phi)] \EE for $u=even$, we are able to calculate it for
$u=odd$, as we have explained already in \Sec{sec:berry}. As a
consequence, we can use $P_a$   in the same way as Berry et.~al.
use $P(\vec{u}_n|\varphi)$ to estimate the unknown phase
$\varphi$. This is the main result of our paper.

In the quantum phase measurement with entangled states, the
generation and detection processes of entangled states are
required separately. On the other hand, in our scheme, the
multiphoton frequency-selective measurement using classical light
generates and detects an effective NOON state simultaneously,
which makes it easier to implement the scheme experimentally.

Another difference between the two schemes is that the measurement
result of the Mach-Zehnder interferometer, which is $u$ and not
the probability distribution $P(\vec{u}_n|\varphi)$, is restricted
to $u$ equals to even or odd. We note that in our approach, we
measure the excitation rate $R^{(2n-1)}\equiv \frac{d}{dt}
P_a(t|\varphi,\phi)$ , which can assume every number between zero
and the maximal excitation rate given by
$R^{(2n-1)}_{\text{max}}\equiv 8 \Omega_{eff}^2 t$. This gives us
more information about the unknown phase $\varphi$. However, if we
want to follow the phase-measurement scheme described in
\cite{Berry2009} exactly, then an excitation rate larger than half
the maximal rate would correspond to $u=even$, and an excitation
rate smaller than half the maximal rate would correspond to
$u=odd$.

\section{Detection Rate Scaling and Error Estimation \label{sec:scaling}}

\subsection{Detection Rate Scaling}
Now we investigate the scaling of maximal excitation rate in our
scheme compared with the one in \cite{Berry2009}, and obtain the
accuracy in supersensitive phase measurement of our scheme.

The phase-measurement scheme described in \cite{Berry2009} is a
realization of the POVM \BE F(\hat \varphi)\equiv
\ket{\varphi}\bra{\varphi}, \quad \ket{\varphi}\equiv
\frac{1}{N_K}\sum\limits_{j=0}^{N_K}\E^{\I j \hat \varphi}\ket{j}
\EE performed on the state \BE \ket{\psi}\equiv
\frac{1}{(N_K+1)^{M/2}}\left(\sum\limits_{j=0}^{N_K} \E^{\I j
\varphi}\ket{j}\right)^{\otimes M} \EE Here, $N_K$ is given by
$N_K=2^{K+1}-1$. Therefore the number of resources is given by
$N\equiv N_K\cdot M$. If the reference phase $\phi$ is chosen
adaptively and for $M\geq 4$, this phase-measurement algorithm
scales like the Heisenberg limit.

Our proposed scheme is just another realization of the
same POVM and therefore also scales like the Heisenberg limit.

In order to find whether it is better to realize with NOON-states
or classical light, we investigate the resources and problems
needed for our realization of the POVM and compare it to the
NOON-state approach. For the NOON-state realization, we have to
create the NOON states $(\ket{2,0}+\ket{0,2})/\sqrt{2}$,
$(\ket{2^2,0}+\ket{0,2^2})/\sqrt{2}$, $\cdots$,
$(\ket{2^K,0}+\ket{0,2^K})/\sqrt{2}$ which is very difficult, and
two multiphoton detectors with high efficiency.  If the
probability of detecting one photon is given by $\eta$, then the
probability of detecting all photons of the NOON state
$(\ket{n,0}+\ket{0,n})/\sqrt{2}$ is given by $\eta^n$ \cite
{Scully2007, Tsang2007}. As a consequence, the detection rate
decreases exponentially with $n$. As mentioned above, Steuernagel
pointed out \cite{Steuernagel2004} the detection rate of entangled
photons arriving at one point will be even smaller since the
photons are spatially unconstrained. Furthermore, the detectors
$c_0$ and $c_1$ need to detect and discriminate between the states
$\ket{0}$, $\ket{1},\dots$, ${\ket{n}}$.  In \cite{Berry2009} the
authors do not explain how to achieve this task.

For the realization with classical light, we need an atom with
multi-level structure. The maximal excitation rate scales like \BE
R^{2n-1}_{max}= 2 \left|\frac{\Omega_S
\Omega_D}{\Delta_{\pm}^2}\right|^{2n}\left|\frac{2\Delta_{\pm}^2}{\Omega_D}\right|^2
t = 2 \left|\frac{2\Delta_{\pm}^2}{\Omega_D}\right|^2 \eta^n t\EE
with \BE \eta \equiv \left|\frac{\Omega_S
\Omega_D}{\Delta_{\pm}^2}\right|^{2}. \EE As a consequence, the
excitation rate decreases exponentially with $n$ similar to the
photon detection rate of the NOON-state approach. However, since a
large number of photons exist in the classical light, although
spatially unconstrained there are enough photons to arrive at one
point to excite the atom (the detector). Therefore, the spatial
distribution of the photons does not affect the excitation rate in
our case.

\subsection{Error Estimation}
Now we derive in the following possible errors  in our scheme from
resonant higher-order terms and from non-resonant terms in a
general case.

For large $n$, strong Rabi frequencies can be used to improve the
excitation rate. We find a tradeoff in using strong Rabi
frequencies such that resonant higher order terms and non-resonant
terms can not be neglected. First, we obtain for the second-order
resonant term from perturbation theory \cite{Scully2007} (which
corresponds to a five-photon process) as Rabi frequencies increase
\begin{widetext}
\begin{eqnarray}
a^{(2)}(t|\varphi,\phi)&=& (-\frac{i}{\hbar})^5 \int\int\int \int
\int dt^5 \bra{a}\mathnormal{H_{I}(\varphi,\phi,t_{1})}
\mathnormal{H_{I}(\varphi,\phi,t_{2})}\mathnormal{H_{I}(\varphi,\phi,t_{3})}
\mathnormal{H_{I
}(\varphi,\phi,t_{4})}\mathnormal{H_{I}(\varphi,\phi,t_{5})}
\ket{b}\nonumber\\
&\equiv& i\Omega_S^2\Omega_D t \left(
\frac{\E^{2\I\varphi}}{\Delta_{1+}\Delta_{2+}}
+\frac{\E^{2\I\phi}}{\Delta_{1-}\Delta_{2-}}\right) \times (\I
r_1-\I r_2-r_3) \end{eqnarray}
\end{widetext}
with \BE r_1 \equiv \frac{\Omega_S^2 t}{\Delta_{1+}}+
\frac{\Omega_S^2 t}{\Delta_{1-}}, \quad r_2 \equiv
\frac{\Omega_S^2 t}{\Delta_{2+}}+ \frac{\Omega_S^2 t}{\Delta_{2-}}
\EE and
\begin{eqnarray}
r_2 &\equiv& \frac{\Omega_D^2 }{\Delta_{1+}\Delta_{2+}}+ \frac{\Omega_D^2 }{\Delta_{1-}\Delta_{2-}} \nonumber \\
&& + \frac{\Omega_D^2 }{\Delta_{1\pm}(\Delta_{2\mp}\pm
\delta)}+\frac{\Omega_D^2 }{(\Delta_{1\mp}\pm
\delta)\Delta_{2\pm}}.
\end{eqnarray}
Here $\mathnormal{H_{I}(\varphi,\phi,t)}$ is obtained from Eq.
\eqref{eq: HI} by replacing $k_{\pm}x$ with $\varphi$ and $\phi$
respectively. The second-order result is obtained by multiplying
Rabi oscillation factors ($r_j$), which describe additional
two-photon processes between intermediate levels, to the first
order contribution $a^{(1)}(x,t)$ in Eq. \eqref{eq:first-order}.
The Rabi oscillation factors are the tendency of resonant higher
order terms, and therefore they have to be much smaller than unity
for perturbation theory to work. We solve this tradeoff by
choosing opposite one-photon detunings,
$\Delta_{j+}=-\Delta_{j-}$, such that $r_1=r_2=0$ while signal
Rabi frequency $|\Omega_S^2t/\Delta_{\pm}|$ can be large. We
choose $\Omega_D^2\ll |\Delta_{(2j-1)\pm}\Delta_{2j\pm}|$
 to suppress the other Rabi oscillation factor
$r_3$. Therefore, under the above conditions of Rabi frequencies
and one-photon detunings, resonant higher-order terms can be
neglected and the excitation rate can be improved.

We discussed the non-resonant terms of a four-level atomic system
in the previous section and now we extend this to a $2n-$level
atomic system. In general, the non-resonant terms result from an
absorption of $n-k$ photons of frequency $\nu_+$ and $k$ photons
of frequency $\nu_-$. The leading term comes from the exchange of
one photon, and therefore the probability amplitude of the state
$\ket{a}$ is given by
\begin{eqnarray}
a(t|\varphi,\phi)&=& \Omega_{eff}t \left[\E^{\I n\varphi}+\E^{\I n\phi}+ n\frac{\E^{\I\delta t}-1}{\I \delta t}\right.\nonumber \\
&&\left. \times\left(\E^{\I(n-1)\varphi+\I\phi}+ \E^{\I\varphi+\I(n-1)\phi}\right)\right]
\end{eqnarray}
with $\Omega_{eff}\equiv \Omega_S^n\Omega_D^{n-1}/\Delta_{\pm}^{2n-2}$. Providing $n \sin (\delta t)/(\delta t)\ll 1$, the probability to find the atom in the excited state is given by
\begin{eqnarray}
P(t|\varphi,\phi)&= &2 (\Omega_{eff}t)^2\left\{1+\cos[n(\varphi-\phi)] +4\frac{n\sin(\delta t)}{\delta t}\right. \nonumber \\
&&\left.\times\cos[\frac{n}{2}(\varphi-\phi)]\cos[\frac{n-2}{2}(\varphi-\phi)]\right\}.
\end{eqnarray}
Assuming, that the unknown phase $\varphi$ is given by $\varphi
\equiv \pi (\varphi_0+\varphi_1/2+\varphi_2/2^2+\dots)$ with $\varphi_n\in
\{0,1\}$ and by applying the phase-measurement algorithm described
in \cite{Berry2009}, the accuracy of obtaining the supersensitive
phase $\varphi_n$ with a detection process involving $n$ photons
of frequency $\nu_\pm$ is given by \BE
\frac{1}{1+\frac{2n\sin(\delta t)}{\delta t}}< 1. \EE

Despite these possible errors, we think our realization of the phase measurement  is more useful, than the one using NOON states, because we think it is easier to find an atom with the right
level structure than to create NOON states with many photons and to find the appropriate detector.

\section{Discussion and conclusion}

We have shown in this paper that the substitution of NOON states
by classical light in the multiphoton frequency-selective
measurement (as suggested for subwavelength lithography in
\cite{Hemmer2006}) is applicable to the phase-measurement scheme
described in \cite{Berry2009} to obtain an accurate
phase-measurement limit. We have found that our scheme is easier
to implement in two ways compared to that in \cite{Berry2009}. The
first advantage is that the multiphoton process using classical
light in our scheme generates and detects a NOON state at the same
time, while the quantum phase measurement with entangled states
requires the generation and detection of entangled states
separately. The second advantage is that, in our scheme, the
multiphoton absorption rate with classical light is not affected
by the spatial distribution of the photons as that in the scheme
using entangled states. Therefore, we conclude that our scheme
with multiphoton frequency-selective measurement using classical
light provides an alternative and better adaptive phase
measurement method.


\section*{Acknowledgement}
This research is supported by NPRP grant (4-346-1-061) by Qatar National Research Fund (QNRF).


\end{document}